\documentclass[aps,prl,floatfix,superscriptaddress,showpacs,twocolumn]{revtex4}
\usepackage{amsmath}
\usepackage{graphicx}
\usepackage[letterpaper,margin=1in]{geometry}

\begin{document}
\title{On the transmission of light through tilted interference filters}
\author{A. O. Sushkov}
\affiliation{Department of Physics, University of California,
Berkeley, CA 94720-7300}
\author{E. Williams}
\affiliation{Department of Physics, University of California,
Berkeley, CA 94720-7300}
\author{D. Budker}
\email{budker@socrates.berkeley.edu} \affiliation{Department of
Physics, University of California, Berkeley, CA 94720-7300}
\affiliation{Nuclear Science Division, Lawrence Berkeley National
Laboratory, Berkeley CA 94720}

\date{\today}
\begin{abstract}
The transmission characteristics of commercial interference filters
are not quite what we expected from the specifications provided by
manufacturers. The unexpectedly sharp dependences of the
transmission coefficient on the incidence angle, wavelength, and on
the position where the light beam falls on the filter may be
important in the analysis of systematic effects in experiments
incorporating interference filters.
\end{abstract}


\maketitle

Despite its official appearance, this is just an informal write-up
intended to summarize some observations that have surprised us (but
perhaps should not have). The book \cite{Mac2001} provides an
excellent discussion of the theory and practice of interference
filters; however, it does not specifically mention the effects we
observe.

\begin{figure}
\includegraphics[width=3.2 in]{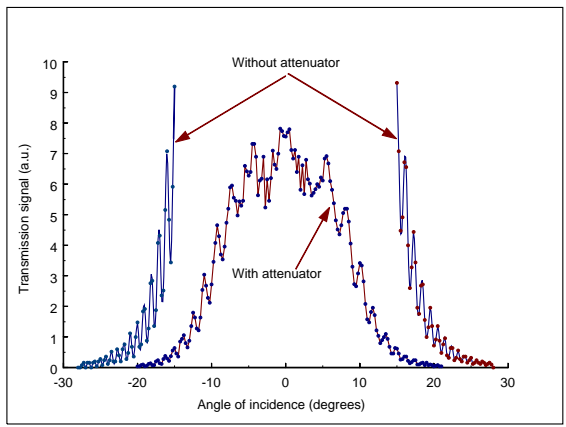}
\caption{Experimental dependence of the transmitted light
intensity as a function of the tilt angle of the filter with
respect to the laser beam. In order to see the details of the
transmission dependence on the angle at large angles where the
transmission through the filter is low, data were taken with the
attenuator removed from the beam path.} \label{FigTrans_v_angle}
\end{figure}
The setup that we used for measuring the transmission
characteristics of interference filters is straightforward. The
light beam from a Spectra Physics Model 117A He-Ne laser
($\lambda=632.8\ $nm) is attenuated with a color-glass plate
(transmission $\approx 10\%$). It then passes through the
interference filter under investigation, which is mounted on a tilt
stage, in turn mounted on a multi-coordinate translation stage. The
transmitted light intensity is measured with a silicon photodiode.

Figure \ref{FigTrans_v_angle} shows the transmission through a 10-nm
bandpass filter with central wavelength $630 $nm (manufacturer
unknown) as a function of the tilt angle of the normal to the filter
with respect to the nominal direction of the laser beam. The colored
plate was positioned $1 $cm away from the laser, the interference
filter was placed $5 $cm from the plate, and the photodiode was 4.5
cm after the filter.  The filter was rotated about an axis
perpendicular to the laser beam and measurements were made at
0.25-degree intervals.

As expected, the transmission falls off as the tilt angle becomes
large. However, instead of the expected smooth dependence, we see
that there is an oscillating structure superimposed on the smooth
dependence with $\sim 10$ periods over the transmission window.
The oscillating structure comprises $\sim 15\%$ near zero tilt
angles, but its relative amplitude increases as the overall
transmission decreases. At large tilt angles, the oscillation
becomes an essentially $100\%$ effect. In order to see the
large-angle regions more clearly, we have also recorded the
transmission with the color-glass attenuator removed (Fig.
\ref{FigTrans_v_angle}).

\begin{figure}
\includegraphics[width=3.2 in]{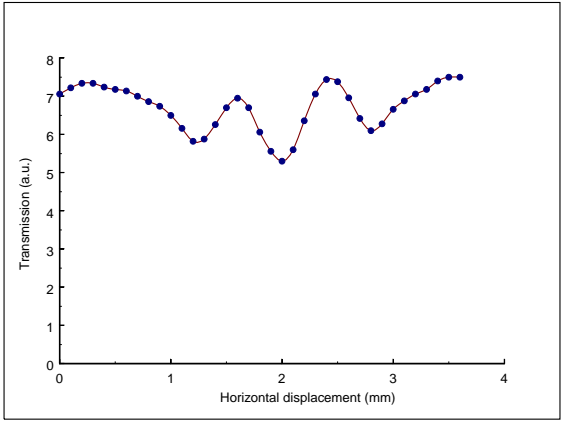}
\caption{An example of the experimental dependence of the
transmitted light intensity as a function of the filter
displacement in the direction perpendicular to the laser beam. The
normal to the filter is tilted by 8 degrees with respect to the
direction of the collimated laser beam (no lens is used for this
measurement).} \label{Fig_Hor}
\end{figure}

We have then looked at the spatial uniformity of the transmission
through the filter with a fixed tilt angle. This was done by
moving the filter with a fixed tilt angle in a direction
perpendicular to the laser beam and monitoring transmission. As
seen in Fig. \ref{Fig_Hor}, we found that, indeed, the
transmission is strongly dependent on the position.

\begin{figure}
\includegraphics[width=3.2 in]{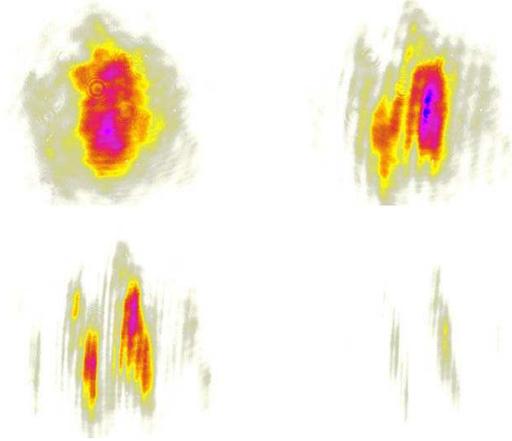}
\caption{Transmission pattern through the same filter for a
diverging laser beam taken with a CCD camera. Different frames
correspond to the tilt angle of the filter by 0 degrees (upper
left frame), 4, 8, and 12 degrees (lower right frame).}
\label{FigIf_Pattern_Unknown}
\end{figure}

\begin{figure}
\includegraphics[width=3.2 in]{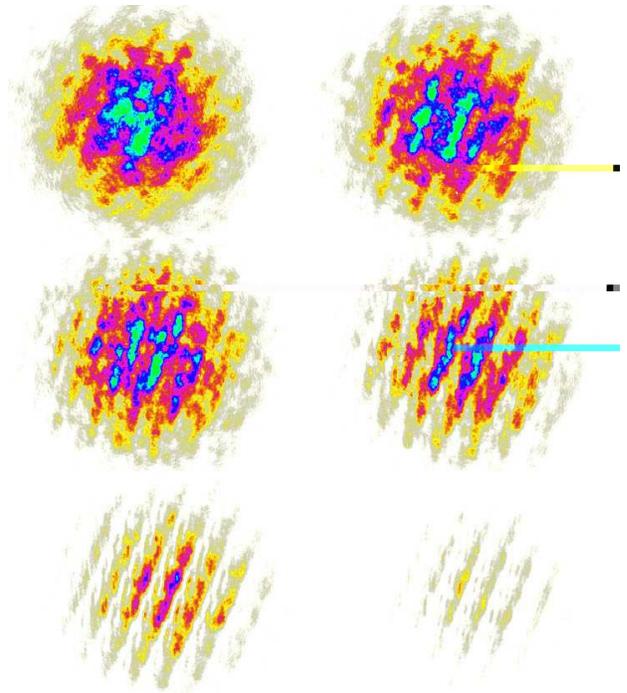}
\caption{Same as in Fig. \ref{FigIf_Pattern_Unknown}, but for
another interference filter (Corion) also centered at 630$\ $nm.
The tilt angles are from 0 to 20 degrees with increment of 4
degrees.} \label{FigIf_Pattern_Corion}
\end{figure}

Next, we examine essentially the same effects, but in a slightly
different arrangement. The images shown in Figs.
\ref{FigIf_Pattern_Unknown} and \ref{FigIf_Pattern_Corion} were
produced by inserting a diverging lens ($f = -6\ $cm) $1\ $cm after
the attenuator (between the attenuator and the interference filter).
The photodiode was replaced with a Coherent LaserCam II CCD camera.
The negative lens converts the 1-mm diameter laser beam to a
diverging beam with a diameter of $1.7\ $mm at the filter and $4.8\
$mm on the camera (with the filter removed).  The interference
filter used to produce the set of images in Fig.
\ref{FigIf_Pattern_Unknown} is the same interference filter as was
used to produce the data of Fig. \ref{FigTrans_v_angle}. The angular
progression of the images shows two families of interference
fringes. The smaller-period fringes are separated by roughly 50
microns while the larger fringe separation is on the order of a
millimeter. Rotating the interference filter about the normal to its
surface going through the point of intersection with the laser beam,
we see that the larger-period structure rotates by the same angle as
the filter, however, the smaller-period fringes do not rotate.
Figure \ref{FigIf_Pattern_Corion} shows images from a 630-nm
interference filter made by Corion. Again, the larger-period fringes
rotated with the filter, while the smaller-period fringes did not.
We have checked (by placing a linear polarizer in front of the
interference filter and rotating it around) that the transmission
pattern was not visibly affected by the direction of the light
polarization.


The effects described above seem to be rather universal. We have
observed interference fringes qualitatively similar to the ones
discussed above with several filters at different wavelengths with
dye lasers in the yellow-red and an Ag hollow-cathode lamp
emitting narrow resonance lines at 328 and 338 nm as light
sources.

Obviously, it is important for anyone who is working with
interference filters and narrow-band light to be aware of the
transmission peculiarities such as the ones described in this
note. Clearly, the oscillations such as the ones shown in Fig.
\ref{FigTrans_v_angle} would also show up in a plot of
transmission vs. the wavelength for monochromatic light. Yet, when
a transmission plot arrives with a newly-purchased filter (or when
we take the transmission curve ourselves using a scanning
spectrophotometer), oscillations are usually not there. Probably,
this is because the spectral window of the spectrophotometer is
broader than the characteristic period of the oscillation, so the
oscillations average in the recorded spectrum.

Of course, it is not too surprising to see interference effects,
albeit unwanted ones, in an interference filter. It would be
interesting to apply thin-film-simulation software to model a
specific filter to figure out the exact origin of the observed
effects (and, perhaps, figure out ways to eliminate them). The
rotation of an interference pattern with rotation of the filter
suggests that this pattern is related to wedging -- imperfect
parallelism between layers of the filters.

We are grateful to A. T. Nguyen and D. English for their help with
testing filter transmission, and to V. V. Yashchuk for very
helpful discussions.

\end{document}